\documentclass[floats,aps,showpacs,twocolumn,superscriptaddress]{revtex4-1}

\usepackage[dvips]{graphicx}
\usepackage{amsfonts}
\usepackage{amsmath,amssymb}
\usepackage{dsfont}
\usepackage{placeins}
\usepackage{afterpage}
\usepackage{flafter}
\usepackage{color}

\newcommand{\myh}{h_{\text{eff}}}     % heff
\newcommand{\myexp}{\beta}            % Power-law exponent
\newcommand{\N}{N}                    % number of states
\newcommand{\Nreg}{\N_{\reg}}         % number of regular states
\newcommand{\Nch}{\N_{\ch}}           % number of chaotic states
\newcommand{\dos}[1]{\rho_{#1}}       % density of states

\newcommand{\Ps}[2]{P_{#2}(#1)}                     % spacing distribution
\newcommand{\ps}[2]{p_{#2}(#1)}                     % spacing contribution
\newcommand{\pszero}[2]{p^{(0)}_{#2}(#1)}           % spacing contribution to 
\renewcommand{\exp}[1]{\text{exp}\left( #1 \right)} % modified exponential
\newcommand{\expalt}[1]{\text{e}^{#1}}              % alternative exponential
\newcommand{\ket}[1]{\left|#1\right>}               % ket state
\newcommand{\X}[2]{X_{#2}\left(#1\right)}           % X function

\newcommand{\reg}{\text{r}}      % regular index
\newcommand{\ch}{\text{c}}       % chaotic index
\newcommand{\cc}{\ch-\ch}        % index for chaotic to chaotic spacings
\newcommand{\rr}{\reg-\reg}      % index for regular to regular spacings
\newcommand{\mixed}{\reg-\ch}    % index for regular to chaotic spacings
\newcommand{\textcc}{c-c }       % index for chaotic to chaotic spacings in txt
\newcommand{\textrr}{r-r }       % index for regular to regular spacings in txt
\newcommand{\textmixed}{r-c }    % index for regular to chaotic spacings in txt
\newcommand{\ireg}{m}            % index of a special regular state
\newcommand{\jch}{j}             % index of a special chaotic state

\newcommand{\mye}{\varepsilon}    % level
\newcommand{\myereg}{\mye_{\reg}} % regular level
\newcommand{\myech}{\mye_{\ch}}   % chaotic level
\newcommand{\mys}{s}              % spacings
\newcommand{\myst}{\tilde{s}}     % spacings
\newcommand{\vt}[1]{v_{#1}}             % typical couplings
\newcommand{\vtilde}[1]{\tilde{v}_{#1}} % scaled couplings

\begin{document}

\title{Fractional-Power-Law Level-Statistics due to Dynamical Tunneling}

\author{Arnd B\"acker}\
\affiliation{Institut f\"ur Theoretische Physik, Technische Universit\"at
             Dresden, 01062 Dresden, Germany}
\affiliation{Max-Planck-Institut f\"ur Physik komplexer Systeme, N\"othnitzer
Stra\ss{}e 38, 01187 Dresden, Germany}

\author{Roland Ketzmerick}\
\affiliation{Institut f\"ur Theoretische Physik, Technische Universit\"at
             Dresden, 01062 Dresden, Germany}
\affiliation{Max-Planck-Institut f\"ur Physik komplexer Systeme, N\"othnitzer
Stra\ss{}e 38, 01187 Dresden, Germany}

\author{Steffen L\"ock}
\affiliation{Institut f\"ur Theoretische Physik, Technische Universit\"at
             Dresden, 01062 Dresden, Germany}

\author{Normann Mertig}
\affiliation{Institut f\"ur Theoretische Physik, Technische Universit\"at
             Dresden, 01062 Dresden, Germany}
\date{\today}

\begin{abstract}
For systems with a mixed phase space we demonstrate that dynamical tunneling
universally leads to a fractional power law of the level-spacing distribution
$P(s)$ over a wide range of small spacings $s$. Going beyond Berry-Robnik
statistics, we take into account that dynamical tunneling rates between the
regular and the chaotic region vary over many orders of magnitude. This results
in a prediction of $\Ps{\mys}{}$ which excellently describes the spectral data
of the standard map. Moreover, we show that the power-law exponent is
proportional to the effective Planck constant $\myh$.
\end{abstract}
\pacs{05.45.Mt, 03.65.Sq, 03.65.Xp, 05.45.Pq}

\maketitle
\noindent

\enlargethispage{\baselineskip}

Spectra of quantum systems whose classical dynamics are either regular or
chaotic usually show universal statistical properties. This fascinating relation
between classical motion and quantum spectra is demonstrated in
Ref.~\cite{BerTab1977}, where it was argued that the spectra of generic regular
systems show Poissonian statistics. In contrast, spectral correlations of
classically chaotic systems can be described by random matrix theory
\cite{BohGiaSch1984, CasValGua1980}. A justification of this conjecture
was given in terms of periodic orbit theory \cite{Ber1985,SieRic2001,Heu2007}. 
The nearest-neighbor level-spacing distribution $\Ps{\mys}{}$ is of central
importance to the study of universal spectral properties \cite{Meh2004, Haa2001,
Sto1999}. Results from these studies are of broad interest for applications in,
e.g.\ solid state physics \cite{ZhoCheZhaYuLuShe2010}, mesoscopic physics
\cite{LibStaBur2009}, cold atom physics \cite{KolBuc2004}, and atomic as well as
acoustic physics \cite{GuhMueWei1998}.

\begin{figure}[b]
  \begin{center}
    \includegraphics[angle = 0]{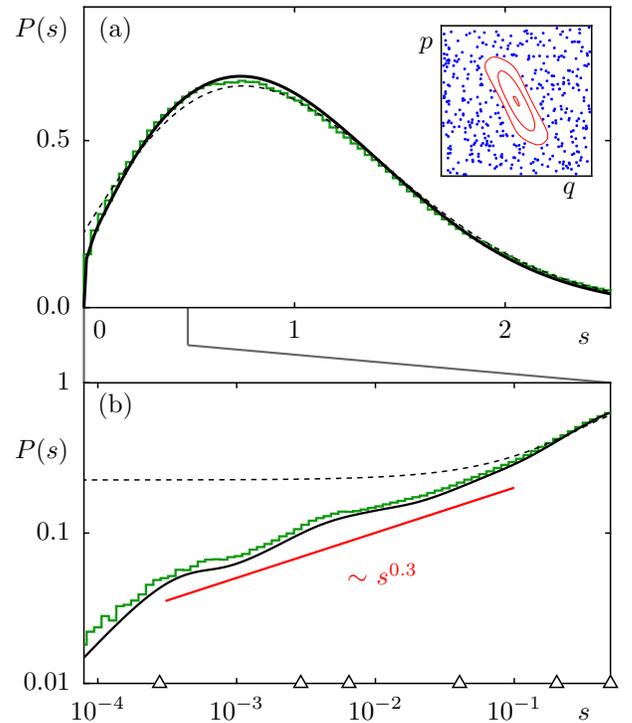}
    \caption{(color online) Level-spacing distribution $\Ps{\mys}{}$ at
$\myh=1/100$ of the standard map \cite{Mapstandard}, with the classical phase
space shown in the inset. We compare the numerical data [(green) histogram] to
the analytical prediction, Eq.~\eqref{Ps}, (black solid line) and the
Berry-Robnik result \cite{BerRob1984} (black dashed line) (a) on a linear and
(b) on a double-logarithmic scale. The typical couplings $2\vt{\ireg}$
(triangles) mark the power-law regime [straight (red) line]. Below the smallest
typical coupling $2\vt{0}$ one finds linear level repulsion.}
    \label{fig1}
  \end{center}
\vspace*{-0.5cm}
\end{figure}

The spacing distribution of generic Hamiltonian systems has been the subject of
an active debate over the last decades \cite{BerRob1984, ProRob1993a,
ProRob1994,Rel2008, AbuDieFriRic2008, VidStoRobKuhHoeGro2007, BatRob2010,
PodNar2003, PodNar2007}. These systems show a mixed phase space, where disjoint
regions of either regular or chaotic motion coexist [see Fig.~\ref{fig1}].
Assuming statistically independent subspectra corresponding to regular and
chaotic regions in phase space, Berry and Robnik computed the level-spacing
distribution of mixed systems \cite{BerRob1984}. In contrast to the predicted
level-clustering behavior, $\Ps{\mys}{}>0$ at $\mys=0$, numerically a fractional
power-law distribution
\begin{eqnarray}
	\label{Power-law}
	\Ps{\mys}{} \propto \mys^{\beta}
\end{eqnarray}
for small spacings $\mys$ with exponent $\myexp\in[0,1]$ was observed
\cite{ProRob1993a, ProRob1994}. Qualitatively this behavior may be described by
the Brody distribution \cite{Bro1973, Izr1990}, as recently discussed in
Refs.~\cite{ZhoCheZhaYuLuShe2010, LibStaBur2009}. Yet, this approach involves a
free fitting parameter which is not related to any physical property of the
system.

Dynamical tunneling \cite{DavHel1981, PodNar2003, BaeKetLoeSch2008,
LoeBaeKetSch2010, BaeKetLoe2010,BohTomUll1993, TomUll1994} plays an important
role for the level-spacing distribution, as it weakly couples regular and
chaotic states and thus enlarges small spacings between the corresponding
levels. In Refs.~\cite{Rel2008, AbuDieFriRic2008, VidStoRobKuhHoeGro2007,
BatRob2010} a phenomenological coupling strength between regular and chaotic
states was introduced, while in Refs.~\cite{PodNar2003,PodNar2007} a fit-free
prediction of the level-spacing distribution was given. However, these results
do not explain the numerically observed power-law distribution,
Eq.~\eqref{Power-law}.

\enlargethispage{\baselineskip}

\begin{figure}[tb]
  \begin{center}
    \includegraphics[angle = 0]{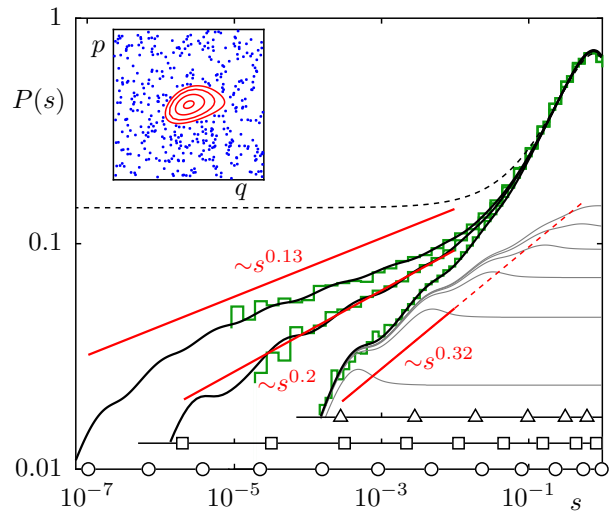}
    \caption{(color online) Comparison of the level-spacing distribution
$\Ps{\mys}{}$ for
the kicked system \cite{Qmapamphib}, with the classical phase space shown in the
inset. The numerical data [(green) histograms], the analytical prediction,
Eq.~\eqref{Ps}, (black solid lines), and the Berry-Robnik result
\cite{BerRob1984} (black dashed line) are shown for $\myh=1/80$ (lower curves),
$\myh=1/120$ (middle curves) and $\myh=1/160$ (upper curves). The corresponding
typical couplings $2\vt{\ireg}$ (triangles, squares, and circles) mark the
power-law regime [straight (red) lines]. The gray solid lines illustrate
contributions from increasing $\vt{\ireg}$ by incrementing the upper summation
index in Eq.~\eqref{Psmixedtunneling} from $0$ to $\Nreg-1$ for $\myh=1/80$.
This gives rise to a staircase function.}
    \label{fig2}
  \end{center}
\end{figure}

In this Letter we show that a fractional power-law distribution universally
arises over a wide range of small spacings because tunneling rates from
different regular states range over many orders of magnitude. We give an
analytical prediction of the level-spacing distribution, which is in excellent
agreement with numerical data of a modified standard map [see Fig.~\ref{fig1}]
and a designed kicked system [see Fig.~\ref{fig2}]. Moreover, we demonstrate
that the power-law exponent $\myexp$ scales like $\beta \propto \myh$. For the
smallest spacings below the power-law regime our result recovers the well known
linear level repulsion \cite{Haa2001}.

As model systems we study kicked systems described by the Hamiltonian
$\tilde{H}(q,p) = T(p) + V(q)\sum_{n\in\mathbb{Z}}\delta(t-n)$
\cite{BerBalTabVor1979}. By a stroboscopic view at integer times one gets an
area preserving map  on the two-torus. The relative areas of the regular and
chaotic regions in phase space are denoted by $\dos{\reg}$ and $\dos{\ch} =
1-\dos{\reg}$. The map is quantum mechanically given by a unitary operator $U$
on a Hilbert space of dimension $\N$ with effective Planck constant $\myh=1/\N$.
The semiclassical limit is approached as $\myh\to 0$. Solving the eigenvalue
equation $U\ket{n}{} = \expalt{\text{i}\phi_{n}}\ket{n}{}$ yields $\N$
eigenphases $\phi_{n}$ with eigenvectors $\ket{n}{}$. According to
the semiclassical eigenfunction hypothesis \cite{Per1973,Ber1977b,Vor1979} one
expects $\Nreg \approx \dos{\reg}\N$ regular states as well as $\Nch \approx
\dos{\ch}\N$ chaotic states. The unfolded spacings are $\mys_{n} =
(\phi_{n+1}-\phi_{n})N/(2\pi)$ for phases $\phi_{n}$ which are
ordered by increasing size.

In order to study the influence of dynamical tunneling on spectral statistics we
consider the standard map \cite{Chi1979} which is the paradigmatic kicked
system. We use parameter values for which it has one large regular island. To
obtain significant statistics for small spacings we consider an ensemble of
modified standard maps, where the regular island is preserved, while the chaotic
dynamics is varied. Thus, the regular levels and the tunneling rates remain
essentially unchanged, while the chaotic levels are strongly varied. This is
realized by varying the kicking potential $V(q)$ in the chaotic part
\cite{Mapstandard}. Moreover, since partial barriers in the chaotic region may
affect spectral statistics beyond dynamical tunneling \cite{BohTomUll1993}, we
remove their influence by the above choice of the kicking potential.
Numerically, we focus on the $\myh$-regime with few regular levels (e.g.\
$\Nreg=6$), as semiclassically ($\Nreg \to \infty$) the influence of tunneling
on spectral statistics becomes less pronounced. We find a power-law distribution
for $\Ps{\mys}{}$ at small $\mys$ [see Fig.~\ref{fig1}]. This behavior is also
observed for another ensemble of kicked systems \cite{Qmapamphib} [see
Fig.~\ref{fig2}].

We model the spectral statistics of systems with a mixed phase space by
considering a random matrix Hamiltonian $H$ which contains  regular levels
$\myereg$ and chaotic levels $\myech$ on its diagonal. These levels are coupled
by off-diagonal elements $\vt{\ireg\text{,}\jch}$, which account for tunneling
contributions between the $\ireg$th regular and the $\jch$th chaotic state. $H$
is scaled such that the mean level spacing is unity and can be chosen real and
symmetric for time-reversal invariant systems. The regular levels $\myereg$ are
semiclassically determined by the torus structure of the regular island
\cite{BerBalTabVor1979}. Since $\Nreg$ is small, they do not show the Poissonian
behavior assumed in Refs.~\cite{PodNar2007, VidStoRobKuhHoeGro2007,BatRob2010}.
The chaotic levels $\myech$ behave like eigenphases of a random matrix from the
circular orthorgonal ensemble \cite{Meh2004}. Here we assume that there are no
additional phase-space structures within the chaotic region. The coupling matrix
elements $\vt{\ireg\text{,}\jch}$ are modeled by independent Gaussian random
variables with zero mean. The standard deviation $\vt{\ireg}$ of
$\vt{\ireg\text{,}\jch}$ does not depend on the chaotic state $\jch$ but is
specific for each regular state $\ireg$. In particular $\vt{\ireg}$ is smaller
for states $\ireg$ which quantize closer to the center of the regular island,
$\vt{0}<...<\vt{\Nreg-1}$. The typical coupling $\vt\ireg$ is related to the
tunneling rate $\gamma_{\ireg}$ of the $\ireg$th regular state by
\begin{eqnarray}
	\label{typical-coupling}
	\vt{\ireg} = \frac{\N}{2\pi}\sqrt{\frac{\gamma_{\ireg}}{\Nch}},
\end{eqnarray}
which follows from the dimensionless form of Fermi's golden rule in kicked
systems \cite{BaeKetLoe2010Fermi}. Hence, we model the probability density
$\bar{P}{\left(\vt{}\right)}$ of all couplings by
\begin{eqnarray}
	\label{couplingdensity}
	\bar{P}{\left(\vt{}\right)} = \frac{1}{\Nreg}\sum_{\ireg=0}^{\Nreg-1}
\frac{1}{\sqrt{2\pi}\vt{\ireg}} \expalt{ -\frac{\vt{}^{2}}{2\vt{\ireg}^{2}}}.
\end{eqnarray}
The tunneling rates $\gamma_{\ireg}$ are parameters of the random matrix model
which can either be determined numerically or analytically, e.g.\ using the
fictitious integrable system approach \cite{BaeKetLoeSch2008,BaeKetLoe2010}.
Since the
tunneling rates $\gamma_{\ireg}$ vary over many orders of magnitude, the typical
couplings $\vt{\ireg}$ embrace a wide range on a logarithmic scale [see the
triangles in Fig.~\ref{fig1}]. Hence, in contrast to previous studies
\cite{VidStoRobKuhHoeGro2007, PodNar2007,BatRob2010}
$\bar{P}{\left(\vt{}\right)}$ is not Gaussian but strongly peaked around small
couplings.

In the spirit of the semiclassical eigenfunction hypothesis
\cite{Per1973,Ber1977b, Vor1979} we partition the level-spacing distribution
into three distinct contributions
\begin{eqnarray}
	\label{Ps}
	\Ps{\mys}{} = \ps{\mys}{\rr} + \ps{\mys}{\cc} + \ps{\mys}{\mixed}.
\end{eqnarray}
Here, $\ps{\mys}{\rr}$ describes the fraction of \textrr spacings formed by two
regular levels, $\ps{\mys}{\cc}$ the fraction of \textcc spacings formed by two
chaotic levels, and $\ps{\mys}{\mixed}$ the fraction of \textmixed spacings
formed by one regular and one chaotic level in the superposed
spectrum \cite{BerRob1984,PodNar2007,BatRob2010}.

We evaluate the three contributions by making the following assumptions: (i) The
spacings of the chaotic subspectrum of $H$ can be approximated by the Wigner
distribution $\Ps{\mys}{\ch} = \pi\mys\dos{\ch}^{2}/2
\expalt{-\pi(\mys\dos{\ch})^{2}/4}$ with mean spacing $1/\dos{\ch}$
\cite{BohGiaSch1984,Haa2001}; (ii) consecutive regular levels are separated on
scales larger than the mean level spacing. This is generically the case if there
are less regular than chaotic states ($\dos{\reg}<\dos{\ch}$) and $\myh$ is not
much smaller than the regular region $\dos{\reg}$ ($\myh\lesssim\dos{\reg}$).
Then one has just few regular levels, $\Nreg \approx \dos{\reg}/\myh$, which are
semiclassically determined by the torus structure of the regular island. The
interval between such consecutive regular levels then typically contains chaotic
levels.

The contribution of zeroth order \textmixed spacings $\myst = |\myereg-\myech|$
to the level-spacing distribution, neglecting couplings between regular and
chaotic states, is given by
\begin{eqnarray}
	\label{Psmixedzero}
	 \pszero{\myst}{\mixed} =
2\dos{\reg}\dos{\ch}\,\exp{\frac{-\pi(\myst\dos{\ch})^{2}}{4}}.
\end{eqnarray}
Here $\dos{\ch}$ is the probability to have a chaotic level $\myech$ in the
distance $\myst$ from the regular level $\myereg$, $\int_{\myst}^{\infty}
\Ps{\mys}{\ch}\text{d}\mys = \exp{-\pi(\myst\dos{\ch})^{2}/4}$ is the
probability to have no further chaotic level between $\myereg$ and $\myech$, and
$2\dos{\reg}$ is the probability of a zeroth order \textmixed spacing
to contribute to $\Ps{\mys}{}$ \cite{BerRob1984}.

Dynamical tunneling leads to enlarged \textmixed spacings, which can be modeled
by the $2\times2$ submatrices of $H$
\begin{eqnarray}
	\label{degenaratesubblock}
	\begin{pmatrix}
	\myereg & \vt{} \\
	 \vt{}  & \myech\\
	\end{pmatrix}.
\end{eqnarray}
This relies on degenerate perturbation theory \cite{PodNar2007,
VidStoRobKuhHoeGro2007} and is applicable because typically both \textrr and
\textcc spacings are large compared to the couplings ($\vt{\ireg}\ll1/\dos{\reg
,\,\ch}$). From Eq.~\eqref{degenaratesubblock} we calculate the tunneling
improved \textmixed spacings  $s = \sqrt{\myst^{2} + 4\vt{}^{2}}$ such that
\begin{eqnarray}
	\label{Psmixedinit}
	\hspace*{-0.15cm}\ps{\mys}{\mixed} =
\hspace*{-0.18cm}\int\limits_{-\infty}^{\infty}\hspace*{-0.1cm}\text{d}\vt{}\,
\bar{P}{\left(\vt{}\right)}
\hspace*{-0.05cm}\int\limits_{0}^{\infty}\hspace*{-0.08cm}\text{d}\myst
\,\pszero{\myst}{\mixed} \,\delta\left(\mys -
\sqrt{\myst^{2}+4\vt{}^{2}}\right).\;
\end{eqnarray}
This expression reflects that \textmixed spacings result from all possible
zeroth order \textmixed spacings $\myst$ and all possible couplings $v$, which
are described by $\bar{P}{\left(\vt{}\right)}$.

In order to compute the integrals in Eq.~\eqref{Psmixedinit} we introduce polar
coordinates $(\mys',\varphi)$ with $\myst = \mys'\cos{\varphi}$ and $2\vt{} =
\mys'\sin{\varphi}$ \cite{VidStoRobKuhHoeGro2007}, such that $\ps{\mys}{\mixed}
= \mys \int_{0}^{\pi/2} \text{d}\varphi \,
\bar{P}{\left(\frac{\mys}{2}\sin\varphi\right)}
\, \pszero{\mys\cos\varphi}{\mixed }$. Calculating the remaining integral gives
\begin{eqnarray}
	\label{Psmixed}
	\ps{\mys}{\mixed} =
\pszero{\mys}{\mixed}\frac{1}{\Nreg}\sum_{\ireg=0}^{\Nreg-1}\,\frac{\vtilde{
\ireg}}{\vt{\ireg}}\X{\frac{\mys}{2\vtilde{\ireg}}}{}
\end{eqnarray}
with $\X{x}{} = \sqrt{\pi /2} x \expalt{-x^{2}/4} I_{0}\left(x^2/4\right)$,
where $I_{0}$ is the zeroth order modified Bessel function of the first kind and
$\vtilde{\ireg}
= \vt{\ireg}/\left(1-2\pi\dos{\ch}^{2}\vt{\ireg}^{2}\right)^{1/2}$.

The contribution of \textrr spacings to the level-spacing distribution is
insignificant, due to assumption (ii)
\begin{eqnarray}
	\label{Psreg}
	 \ps{\mys}{\rr} = 0.
\end{eqnarray}

\enlargethispage{\baselineskip}

The contribution of \textcc spacings to $\Ps{\mys}{}$ is
\begin{eqnarray}
	\label{Psch}
	 \ps{\mys}{\cc} = \Ps{\mys}{\ch}\left[1-\dos{\reg}\mys\right],
\end{eqnarray}
where the first factor is the probability of finding a \textcc spacing of size
$\mys$ in the chaotic subspectrum of $H$. The second factor describes the
probability of having no regular level within this \textcc spacing, which is
valid
in the regime where $\mys$ is smaller than all \textrr spacings.

Combining Eqs.~\eqref{Psmixed}-\eqref{Psch} in Eq.~\eqref{Ps} gives our
prediction of the level-spacing distribution in the presence of dynamical
tunneling. Figures~\ref{fig1} and \ref{fig2} show that this
result is in excellent agreement with spectral data of our
example systems for $\mys\lesssim 1$. The small deviations for $\mys\gtrsim 1$
can be attributed to approximation~(i).

\enlargethispage{\baselineskip}

Now we derive the fractional power-law of $\Ps{\mys}{}$, Eq.~\eqref{Power-law},
in the tunneling regime. Here $\mys$ is in between the smallest typical coupling
$\vt{0}$ and the largest typical coupling $\vt{\Nreg-1}$. In this regime \textrr
spacings do not contribute [see Eq.~\eqref{Psreg}]. Furthermore, the repulsion
between chaotic levels [see Eq.~\eqref{Psch}] ensures that \textcc spacings only
have significant probability for larger spacings and are insignificant deep in
the tunneling regime ($\mys\ll 1$). Here, Eq.~\eqref{Psmixedzero} reduces to
$\pszero{\mys}{\mixed} \approx 2\dos{\reg}\dos{\ch}$ which is used together with
$\vtilde{\ireg} \approx \vt{\ireg}$ in Eq.~\eqref{Psmixed} leading to
\begin{eqnarray}
	\label{Psmixedtunneling}
	\Ps{\mys}{} \approx
\frac{2\dos{\reg}\dos{\ch}}{\Nreg}\sum_{\ireg=0}^{\Nreg-1}\X{\frac{\mys}{2\vt{
\ireg}}}{}.
\end{eqnarray}
$\X{\mys/2\vt{\ireg}}{}$ behaves linearly for $\mys < 2\vt{\ireg}$ and reaches a
plateau of unit height for $\mys > 2\vt{\ireg}$. Hence, one finds linear level
repulsion below the smallest coupling $\vt{0}$ which would dominate the whole
tunneling regime, if only one typical coupling was present in the system
\cite{VidStoRobKuhHoeGro2007,PodNar2007,BatRob2010}. However, the $\vt{\ireg}$
range over many orders of magnitude. This leads to an arrangement of the $X$
functions, which individually behave like a step function, such that
$\Ps{\mys}{}$ becomes a staircase on a double logarithmic scale [see
Fig.~\ref{fig2}]. This staircase resembles a power law, according to
Eq.~\eqref{Power-law}, and its slope is the power-law exponent
\begin{eqnarray}
  \label{exponent}
  \myexp \approx \log{(\Nreg)}/\log{(\vt{\Nreg - 1}/\vt{\text{0}})}.
\end{eqnarray}
This explicitly shows that the fractional power-law arises from typical
couplings which range over many orders of magnitude.

We now evaluate the scaling of the exponent $\myexp$ under variations
of $\myh$. With $\Nreg = \dos{\reg}/\myh$, $\vt{ \small{\Nreg} - 1 } \approx 1$,
and the rough estimate $\vt{\text{0}} \sim \exp{-C\dos{\reg}/\myh}$
\cite{HanOttAnt84,BaeKetLoeSch2008}, we get
\begin{eqnarray}
	\label{exponentscaling}
	\myexp \propto \myh.
\end{eqnarray}
This type of scaling behavior, phenomenologically derived in
Ref.~\cite{ProRob1994}, is confirmed in Fig.~\ref{fig2} ($\myh = 1/80$, $1/120$,
and $1/160$), with the almost constant ratio $\myexp/\myh =$ 26, 24, and 21 for
the three cases of $\myh$. This result demonstrates the inapplicability of the
Brody distribution \cite{Bro1973,Izr1990} for mixed systems, as it fails to
simultaneously describe the $\myh$-dependent power-law exponent for small
spacings and the fixed Berry-Robnik-type distribution of large spacings beyond
the tunneling regime.

Let us conclude by considering the level-spacing distribution in the
semiclassical limit $\myh\to 0$. In this limit small \textrr spacings appear
such that assumption (ii) is violated. A generalized derivation shows that the
\textmixed contribution to $\Ps{\mys}{}$ then still follows a power-law with
exponent $\myexp\to 0$, as in Eq.~\eqref{exponentscaling}. At the same time the
\textrr contribution approaches Poisson statistics. In combination one thus
recovers Berry-Robnik statistics in the semiclassical limit.

To summarize, we have demonstrated how the wide range of dynamical tunneling
rates universally leads to a power-law distribution of $\Ps{\mys}{}$ at small
spacings $\mys$. We expect that this is also the fundamental mechanism which
explains spacing-statistics in mixed systems with more complicated phase-space
structures.

We thank the DFG for support within the Forschergruppe 760 "Scattering Systems
with Complex Dynamics".


\newpage

\begin{thebibliography}{10}


\bibitem{BerTab1977}
M.~V.~Berry and M.~Tabor, Proc. R. Soc. Lond. {\bf 356}, 375 (1977).

\bibitem{BohGiaSch1984}
O.~Bohigas, M.~J.~Giannoni, and C.~Schmit, Phys. Rev. Lett. \textbf{52}, 1
(1984).

\bibitem{CasValGua1980}
G.~Casati, F.~Valz-Gris, and I.~Guarnieri, Lett. Nuovo Cimento \textbf{28}, 279
(1980).

\bibitem{Ber1985}
M.~V.~Berry, Proc. R. Soc. Lond. \textbf{400}, 229 (1985).

\bibitem{SieRic2001}
M.~Sieber and K.~Richter, Physica Scripta \textbf{T90}, 128 (2001).

\bibitem{Heu2007}
S.~Heusler, S.~M\"uller, A.~Altland, P.~Braun, and F.~Haake, Phys. Rev. Lett.
\textbf{98}, 044103 (2007).

\bibitem{Meh2004}
M.~L.~Mehta, \emph{Random Matrices}, (Elsevier, Boston, 2004).

\bibitem{Haa2001}
F.~Haake, \emph{Quantum Signatures of Chaos}, (Springer, Berlin, 2001).

\bibitem{Sto1999}
H.~J.~St\"ockmann, \emph{Quantum Chaos} (Cambridge University Press, Cambridge,
1999).

\bibitem{ZhoCheZhaYuLuShe2010}
W.~Zhou, Z.~Chen, B.~Zhang, C.~H.~Yu, W.~Lu, and S.~C.~Shen, Phys. Rev. Lett.
\textbf{105}, 024101 (2010).

\bibitem{LibStaBur2009}
F.~Libisch, C.~Stampfer, and J.~Burgd\"orfer, Phys. Rev. B \textbf{79}, 115423
(2009).

\bibitem{KolBuc2004}
A.~R.~Kolovsky and A.~Buchleitner, Europhys. Lett. \textbf{68}, 632 (2004).

\bibitem{GuhMueWei1998}
T.~Guhr, A.~M\"uller-Groeling, and H.~A.~Weidenm\"uller, Phys. Rep.
\textbf{299}, 189 (1998).

\bibitem{BerRob1984}
M.~V.~Berry and M.~Robnik, J. Phys. A \textbf{17}, 2413 (1984).

\bibitem{ProRob1993a}
T.~Prosen and M.~Robnik, J. Phys. A \textbf{26}, 2371 (1993).

\bibitem{ProRob1994}
T.~Prosen and M.~Robnik, J. Phys. A \textbf{27}, 8059 (1994).

\bibitem{PodNar2003}
V.~A.~Podolskiy and E.~E.~Narimanov, Phys. Rev. Lett. \textbf{91}, 263601
(2003).

\bibitem{PodNar2007}
V.~A.~Podolskiy and E.~E.~Narimanov, Phys. Lett. A \textbf{362}, 412 (2007).

\bibitem{VidStoRobKuhHoeGro2007}
G.~Vidmar, H.-J. St\"{o}ckmann, M.~Robnik, U.~Kuhl, R.~H\"{o}hmann, and
S.~Grossmann, J. Phys. A \textbf{40}, 13883 (2007).

\bibitem{Rel2008}
A.~Rela\~no, Phys. Rev. Lett. \textbf{100}, 224101 (2008).

\bibitem{AbuDieFriRic2008}
A.~Y.~Abul-Magd, B.~Dietz, T.~Friedrich, and A.~Richter, Phys. Rev. E
\textbf{77}, 046202 (2008).

\bibitem{BatRob2010}
B.~Batisti{\'c} and M.~Robnik, J. Phys. A \textbf{43}, 215101 (2010).

\bibitem{Bro1973}
T.~A.~Brody, Lett. Nuovo Cimento \textbf{7}, 482 (1973).

\bibitem{Izr1990}
F.~M.~Izrailev, Phys. Rep. \textbf{196}, 299 (1990).

\bibitem{DavHel1981}
M.~J.~Davis and E.~J.~Heller, J. Chem. Phys. \textbf{75}, 246 (1981).

\bibitem{BohTomUll1993}
O.~Bohigas, S.~Tomsovic, and D.~Ullmo, Phys. Rep. \textbf{223}, 43 (1993).

\bibitem{TomUll1994}
S.~Tomsovic and D.~Ullmo, Phys. Rev. E \textbf{50}, 145 (1994).

\bibitem{BaeKetLoeSch2008}
A.~B\"{a}cker, R.~Ketzmerick, S.~L\"{o}ck, and L.~Schilling, Phys. Rev. Lett.
\textbf{100}, 104101 (2008).

\bibitem{LoeBaeKetSch2010}
S.~L\"ock, A.~B\"acker, R.~Ketzmerick, and P.~Schlagheck, Phys. Rev. Lett.
\textbf{104}, 114101 (2010).

\bibitem{BaeKetLoe2010}
A.~B\"acker, R.~Ketzmerick, and S.~L\"ock, Phys. Rev. E \textbf{82} 056208
(2010).

\bibitem{Mapstandard}
We consider the standard map on the torus with unit length, using $T(p)=p^{2}/2$
and $V(q)=\kappa\,\cos{(2\pi q)}/(2\pi)^{2}$ for $q\in
[\tilde{q},1-\tilde{q}]$. For $q\in [0,\tilde{q}]$ and $q\in [1-\tilde{q},1]$ we
use the modified potential $V(q)=c(q^{2} - \tilde{q}^{2})/2 + \kappa\cos{(2\pi
\tilde{q})}/(2\pi)^{2}$ and $V(q)=c((q-1)^{2} - (\tilde{q}-1)^{2})/2 +
\kappa\cos{(2\pi \tilde{q})}/(2\pi)^{2}$, respectively. We choose $\kappa=3.0$
and $\tilde{q}=0.275$ such that the regular region (of size
$\dos{\reg}\approx0.12$) of the standard map is preserved. Taking $10^{7}$
equidistant values $c\in [10, 1000]$ enables an ensemble average.

\bibitem{BerBalTabVor1979}
M.~V.~Berry, N.~L.~Balazs, M.~Tabor, and A.~Voros, Ann. Phys.
(N.Y.) \textbf{122}, 26 (1979).

\bibitem{Per1973}
I.~C.~Percival, J. Phys. B \textbf{6}, L229 (1973).

\bibitem{Ber1977b}
M.~V.~Berry, J. Phys. A \textbf{10}, 2083 (1977).

\bibitem{Vor1979}
A.~Voros, in \emph{Stochastic Behavior in Classical and Quantum Hamiltonian
Systems} (Springer-Verlag, Berlin, 1979), no.~93 in Lecture Notes in Physics,
pp. 326--333.

\bibitem{Chi1979}
B.~V.~Chirikov, Phys. Rep. \textbf{52}, 263 (1979).

\bibitem{Qmapamphib}
Similar to Ref.~\cite{SchOttKetDit2001} we design a kicked system by defining
the functions $t'(p)$ and $v'(q)$ with $t'(p)=-p$ for $p\in [-1/4,1/4]$ and
$t'(p)=p$ for $|p|\in (1/4,1/2)$ as well as $v'(q)=c(2q+1)$ for $q\in
[-1/2,-1/4]$, $v'(q)=-2rq + 4Rq^{2}$ for $q\in (-1/4,1/4)$, and
$v'(q)=(c+3R/2)(2q-1)$ for $q\in[1/4,1/2)$. Smoothing the periodically
extended functions with a Gaussian, $G(z) = \exp{-z^{2}/2\varepsilon^{2}}
/\sqrt{2\pi \varepsilon^{2}}$, yields analytic functions $T'(p) =
\int\text{d}z\,G(z)t'(p+z)$ and $V'(q)=\int\text{d}z\,G(z)v'(q+z)$. We choose
$r=0.26$, $R=0.4$, and $\varepsilon=0.001$, leading to $\dos{\reg}\approx0.07$.
Taking $10^5$ equidistant values $c\in [5, 1000]$ enables an ensemble average.

\bibitem{BaeKetLoe2010Fermi}
Compared to Appendix A in Ref.~\cite{BaeKetLoe2010} we rescale the mean level
spacing to unity with the factor $\N/(2\pi)$.

\bibitem{HanOttAnt84}
J.~D.~Hanson, E.~Ott, and T.~M.~Antonsen, Phys. Rev. A \textbf{29}, 819 (1984).

\bibitem{SchOttKetDit2001}
H.~Schanz, M.-F.~Otto, R.~Ketzmerick, and T.~Dittrich, Phys. Rev. Lett.
\textbf{87} 070601 (2001).

\end{thebibliography}
\end{document}